\documentstyle[12pt,twoside,fleqn,espcrc1,floatfig,epsfig]{article}
\title{
    Strangeness in  ultrarelativistic  nucleus-nucleus collisions} 
\author{ H.\ Sorge \address{ Physics  Department,  
 State University of New York at Stony Brook, NY 11794-3800 
  }}

\begin{document}
\initfloatingfigs
\maketitle         

\begin{abstract}
I discuss strangeness production in  nucleus-nucleus reactions at 
ultrarelativistic energies (up to 200~AGeV). In these reactions  matter may be 
created with densities and temperatures in the transition region between 
quark-gluon plasma (QGP) and hadron gas. Strange anti-baryon enhancement at 
200~AGeV and probably even more so at 10 AGeV signals importance of interactions
beyond hadron gas dynamics. The systematics of strangeness production indicates
that energy and baryon density are key variables while the size of the 
production volume plays no visible role. Analysis of strangeness appears useful
to explore thermalization, flow and the  post-equilibrium stage in 
ultrarelativistic nucleus-nucleus collisions.      
\end{abstract}
  
\section{Introduction} 
   
 In this talk I am going to review recent developments to extract
 information on the properties of dense matter 
 from measurements of
 strange particles  
 -- with emphasis on  the phase transition to
  quark-gluon plasma (QGP).      
 Originally,  strangeness enhancement  had 
 been suggested as  a   signature of the quark-gluon plasma (QGP).
 Most notably, strange anti-baryons
 have been viewed as particularly well-suited to monitor the formation
 of a transient QGP in nuclear reactions.
 Strangeness production in  ultrarelativistic  nucleus-nucleus collisions
 has been studied experimentally  for about one decade now, both  at
  the BNL-AGS  and at the CERN-SPS.
  At the AGS  Au beams have been accelerated with an energy of up to
  nearly 12 AGeV. A Pb beam is being utilized at CERN with a beam energy
  of 158 AGeV.
  Indeed, it has been found that strange hadron yields normalized
  to pions are enhanced in $AA$ collisions at these energies.
A strong strange anti-baryon
 enhancement has been experimentally
observed
for central S collisions on S and heavier targets at 200~AGeV 
\cite{NA3590,ABA91}.  An even stronger enhancement 
 of $\bar{\Lambda}/\bar{p}$
 in Si+Au and Au+Au reactions at 10-15~AGeV is  suggested by some
  preliminary data.
 Two  sets of $\bar{p}$ data 
  measured by detectors with different sensitivity to feed-down from
   weak (hyperon) decays
  can be reconciled with each other only if the  $\bar{\Lambda}/\bar{p}$ 
   ratio is extremely large, on the order of 5-6 \cite{HIPAGS96}.
 Over the years it has become clear
 from the systematics of measurements  and  from  theoretical considerations
 that  strangeness enhancement is not as clearly linked  to a QGP as initially
  thought. 
   The theoretical debate about  production mechanisms will probably go 
   on for a while, because calculations  are messy due to the intrinsically
   nonperturbative nature of hadronization. 
   Therefore I  will   discuss later
    whether  one can distinguish  production mechanisms experimentally.
   
  Recently, it has been argued that  central nucleus-nucleus collisions
   at high energies produce  a hadronic state in chemical equilibrium
   at freeze-out  \cite{BMSWX96}. 
   This would be, of course, a radical `solution' to the debate about
 production mechanisms. An equilibrium state does not  keep a memory of how it
   has been produced. The information about its creation would
   be lost.  The situation is not so hopeless as it looks like.           
  By lowering the beam energy or the projectile-target masses one can
  impose more unfavorable  conditions  for chemical equilibration. 
 The final-state dynamics in nuclear reactions adds another facet
 to the thermalization issue.  
 `Freeze-out' in heavy-ion collisions is not a state but a stage.
 It is sometimes argued that only if the final hadronic stage can be
 characterized in terms of thermal and hadrochemical equilibrium one can
 hope to describe the earlier stages in terms of equilibrium concepts
 (see e.g.\ \cite{SBM96}).  
 However, already several years ago
  Bebie et al.\ have demonstrated in a  thorough study of the 
 decoupling conditions in ultrarelativistic $AA$ collisions that the
 various  equilibria  break-down necessarily in the final stages  \cite{BEB92}.
 For instance, 
 strange hadron yields go out of equilibrium first, because strangeness
  production and absorption rates are getting  small in comparison to the
  (quasi-elastic) collision rates. 
  Hydrodynamic and transport calculations  point to the existence of
 a `post-equilibrium' stage   in ultrarelativistic $AA$ reactions.
 One can utilize the effect that different hadron species
   decouple  sequentially to study the time evolution of  the system.
  For instance, this helps to address  the  question  whether the
   observed collective flow develops `early' or `late'.  Such information 
   is critical  for studies of the equation of state in the phase transition
 region \cite{SOR97PLB}.

\section{Prehadronic stage: strings, partons or QGP?} 
   
 Hadronic observables can  probe only  indirectly 
 the nature of the prehadronic stage. Hadronization and
  further interactions in the hadronic phase will  leave their
 imprints on chemical composition and phase space distributions.
  With this caveat 
   strange anti-baryons are probably the most useful species to gain
 information on the  prehadronic stages of nuclear collisions.
 Interactions  in a  hadron gas   are too slow  to result in
  a sizable creation rate.
  This was conjectured first based on calculations
  for a thermal system  \cite{KOC86}. 
  Later, it was confirmed by  transport calculations as well.
  The hadronic interactions in the pre-equilibrium stage are very
  energetic according to  calculations with a transport
  model like RQMD. Initial
  interactions between mesons and in-going baryons are characterized by
 effective temperatures of more than  300 MeV at CERN energy  \cite{SOR95ZFP}.
  Employing $\Delta E \sim 3 T$  tells 
  that $\bar{s}s$ creation  at this stage is not severely suppressed 
   in channels like $\Lambda K$.  However, even these hard initial hadronic
   collisions do  usually not  provide the necessary energy of
   2-4~GeV needed to create   a  strange baryon pair.
   
  Since energy is the key variable it is  natural to
   expect that  the earlier and denser prehadronic stage provides
   more favorable conditions for anti-baryon production.
  The models which have been successfully tested in $hh$, $hA$
   and $AA$ collisions at the ultra-high beam energies 
 $\gg$ 10~AGeV 
 (RQMD , VENUS, QGSM  and other versions of DPM models) 
 employ some sort of 
 multiple string excitation mechanism for the prehadronic (quark matter) stage.
  The strings represent the coherent 
   color-electric field which 
  is radiated  from the in-going 
   color charges. The soft modes of these fields are
    squeezed by the surrounding non-perturbative vacuum into flux-tubes
   essentially along the beam axis.
   If the system is heated quickly
    -- e.g.\  by coupling to hard partonic modes -- 
   string-like configurations may
    `melt' and give  way to a QGP.
  Whether this happens at presently explored beam energies is still an open
  question.
 
\par         
\begin{floatingfigure}{7cm} 

\mbox{
\epsfig{figure=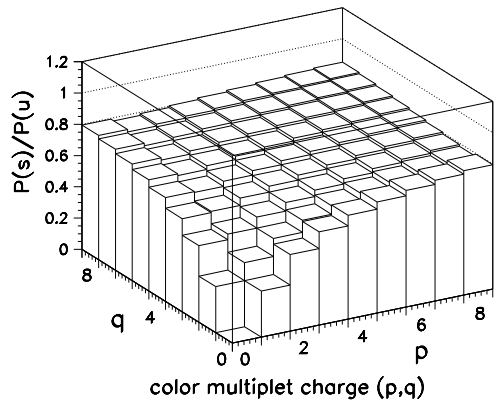,height=6cm,width=6cm}}   
  \caption[
          ]
                    {
 \label{ropessbar}
  {\small
  Rope decay:  \\
  \underline{
 strangeness suppression 
    }}
                            }
 \end{floatingfigure}
   \quad              
  
 Beyond some `critical' string
  density independent fragmentation is expected to break down.
 In RQMD \cite{SOR92} (and also in the Spanish version of the DPM \cite{ABFP95})
  strings will fuse
  into color `ropes'  if they are overlapping.
 One consequence of the strong coherent gluon fields in the rope
 is their fast screening by quark pair production. 
 The strong $s\bar{s}$ enhancement in  the coherent rope fields is
  displayed in Fig.~\ref{ropessbar}.  $(p,q)$ characterize the
  $SU_3$ color charge creating the rope gluon field. $p$, respectively
 $q$ can be thought of as the unscreened quark (anti-quark) sources of
 the field. Unlike in a string, 
 the end of ropes contain  usually many diquarks and anti-diquarks
  with the same   color charge as (anti-)quarks.
  This is the consequence of the $SU_3$ group combinatorics, e.g.\ 
    $3+3=6+\bar{3}$. 
 The most drastic change of rope 
   compared to independent string fragmentation is therefore in the
  strange anti-baryon sector, because strangeness and (anti-)diquark
  enhancement both contribute here. 
  While independent string fragmentation fails completely in 
  comparison to the data for S induced reactions at 200~AGeV
   the results with color rope (or string fusion) included 
  agree rather well with measurements \cite{SOR95ZFP}.
  First preliminary results from NA49  \cite{CBOR97}  and  WA97
  (see K.\ Safarik's contribution to the Proceedings) for central
   Pb(158AGeV) on Pb reactions are consistent with the RQMD
   predictions of anti-baryon densities published in  \cite{SOR95}.
  Anti-baryon rapidity distributions 
 ($\overline{p}$, $\overline{\Lambda }$, 
   $\overline{\Xi}^{\: 0,+}$)
  in central Pb(160AGeV)+Pb collisions
  which have been calculated using RQMD in three different modes
   are displayed in Fig.~\ref{pbpbbbar}.
  The three different operation modes of RQMD  have been
  ropes and rescattering switched off (`NN mode', dashed line),
   rope fragmentation included (dotted line)
 and ropes and hadronic rescattering both included
  which is the default mode (straight line).   
  The   $\overline{\Xi}$ yield as calculated with the default
    version 2.1 of RQMD 
    increases by a factor of 7.7
   compared to the result in the `NN mode'.
    The increase would be 
     even a factor of 13.3 if anti-baryons
   would not be absorbed in the later hadronic stage. 
  
  Capella has taken a somewhat different path to explain strangeness and 
 diquark enhancement from string fragmentation in $AA$ collisions \cite{CAP95}.
   Elementary $hh$ collisions produce mostly strings with
  the in-going valence quarks at their ends. However,   in $AA$ collisions
   additional strings are formed with sea quarks at the ends.
   With an assumption about a rather large  strangeness fraction
    $s/u$=0.5
   and some $B$-$\bar{B}$
  content in the nucleon sea he can explain measurements for 
   S+S and S+W collisions at 200 AGeV.
   It is an interesting question whether the large diquark and strangeness
   content in this model mimicks the effect of rope formation or the
  other way  around. 

\par         
\begin{floatingfigure}{7cm} 

\mbox{
\epsfig{figure=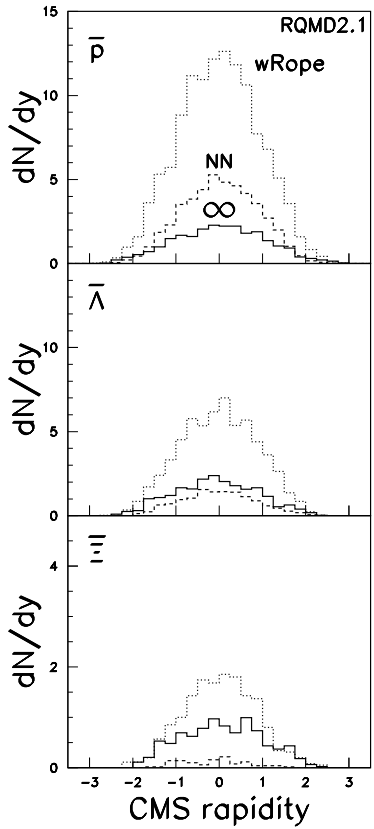,height=12cm,width=6cm}}   
  \caption[
          ]
                    {
 \label{pbpbbbar}
  {\small
  Pb(160AGeV)+Pb,  \\
  \underline{
 Anti-baryons from RQMD
    }}
                            }
\end{floatingfigure}
   \quad              
 A  completely statistical (microcanonical) approach 
 to the fragmentation of large density spots 
 which emerge in a percolation of strings  has been  
 developed in  VENUS \cite{WER93}. The chemical composition depends on the   
 choice for the   energy at which the  `droplet'  is converted into hadrons. 
 VENUS iterates this scheme  to simulate
 hadronic interactions. Since only final hadron yields from VENUS have been
  published it is impossible to directly compare the chemical compositions
   from initial droplet decay and rope fragmentation.
 Qualitatively, trends -- strangeness and even more pronounced anti-baryon
   enhancement  --
 are similar in both approaches. 
 Note that  in RQMD anti-baryon annihilation on baryons is based on
 the free $\bar{p}p$ cross sections and in VENUS on 
  a smaller --  energy-independent and universal  -- interaction strength.

 Very recently, it has been claimed that one can explain soft particle
  production without invoking strings -- by straightforward application
  of perturbative QCD in a parton cascade model (PCM)  \cite{GEI97}.
 However,  a `string effect' is effectively put     
 into  this model 
    using  arbitrary cut-off prescriptions for the 
  infrared-divergent parton cross sections. 
 The cut-offs have been fine-tuned to generate the  rapidity densities which
 have been experimentally observed in $hh$ collisions. 
  The  place in which 
    hard parton-parton scatterings and not  the cut-off's determine the
   model's results are hadrons with  large transverse momenta. 
  Incidentally, PCM  generates  too many
  pions with $p_t > $3 GeV --
   a factor 5 compared to experimental 
   WA98 data. 
   The failure points either to an overestimation
   of multiple hard collision effects in PCM or an underestimation of
   the energy loss which a fast parton experiences in dense matter
   due to lack of soft interactions. 
 The parton cascade 
  combined with a hadronic `after-burner' (HIJET) agrees well with preliminary
 data from NA49 for $K_S$/$\pi$. Interestingly, the strangeness
  enhancement (compared to $pp$)  seems to have been produced  mainly in the
 hadronic stage. 
  A  similar trend has been found in RQMD comparing strangeness enhancement
  in rope fragmentation with the contribution from hadronic rescattering
  \cite{SOR95}.
  Concerning strange anti-baryon production the  parton cascade model 
  shows  a weaker enhancement for central Pb(158AGeV) on Pb
 reactions than  VENUS (and RQMD). This
 can be seen from the model comparisons 
 of $\bar{\Lambda}/\bar{p}$ in  \cite{GEI97}.
  
  Some statistical approaches have been developed in recent years to 
   describe the hadronization of a quark-gluon plasma microscopically,
   mostly in terms of quark coalescence                
   (see e.g.\  \cite{BLZ95},\cite{RLT97}). 
  These coalescence models predict an $\bar{\Omega}/\Omega $ ratio  $\ge$ 1. 
  The result is  generic for this class of models, because 
  there are  as many  $s$  as $\bar{s}$ quarks,
  independent on the degree of strangeness equilibration.
   Furthermore, three
  $s$ quarks fuse with  same  probability  as  three $\bar{s}$ quarks 
  into  an $\Omega$, respectively $\bar{\Omega}$ state. 
  So far, no further interactions between hadrons are included in these 
  models.
     
   The differences in the treatment of  anti-baryon {\em annihilation} 
     in the various quoted approaches are clearly larger than for
     the production mechanisms.  
    It ranges from no annihilation at all to  full  strength as determined
    from free $\bar{p}$$p$ cross sections. 
   Therefore it is useful     
    to explore the strength of  anti-baryon 
    absorption  experimentally.
   For instance, measurement of `anti-flow' 
   in non-central collisions can  clarify how strongly
     various anti-baryon species are absorbed by the co-moving baryons
  \cite{JAH94}.
  Another idea is to study  anti-baryon-proton correlations in the final
   state. 
    Note that anti-baryon studies in the `stopping region' 
    (up to 30 GeV) are particularly useful in this respect.
   Only a small fraction of anti-baryons survives in the extremely
    baryon-rich medium produced at these energies.
   Comparison between models and experimental data has lead to
    quite detailed insights. Most interestingly, the 
     the $\bar{p}p$ annihilation cross section which rises
   strongly  at low energy may be somewhat screened in dense matter
    \cite{ARC93}.  In most recent RQMD versions  this 
    effect is  attributed to  $\bar{B}-B$ `molecule' states 
     close to threshold which have a  finite life time  before
    they radiate into mesons. In the medium they may break up again
   if they scatter with other hadrons.

\section{Equilibration:  is p+A  more interesting than Pb+Pb?} 
   
  As pointed out already as a caveat interactions in the hadronic phase
   may  strongly alter  the primordial anti-baryon distributions. 
   This holds certainly true for anti-baryons in baryon-rich matter,
    because their  annihilation  rate based on free-space cross sections
    is huge.
   We know from experimental data that a sizable fraction of the in-going
   baryons is stopped. Therefore annihilation reactions may  
    possibly wash out  the interesting information from the
   earlier stage. 
    A very  strong effect of anti-baryon absorption in
    central Pb(160AGeV) on Pb collisions  can be inferred from
   the RQMD results in Fig.~\ref{pbpbbbar}.
   Note, in particular,  that the final $\bar{p}$ is even below the
    result from the $NN$ mode of RQMD. All strong enhancement due
   to initial rope formation is gone! This provides  a useful lesson
  about the study  of production mechanisms. Heavy projectile-target
  combinations  with beam energy 
   as large as 160~AGeV are probably unsuitable for this purpose, because
   strong chemical equilibration destroys the memory of the earlier stages. 
     
\begin{figure}[tb]

\centerline{\hbox{
\psfig{figure=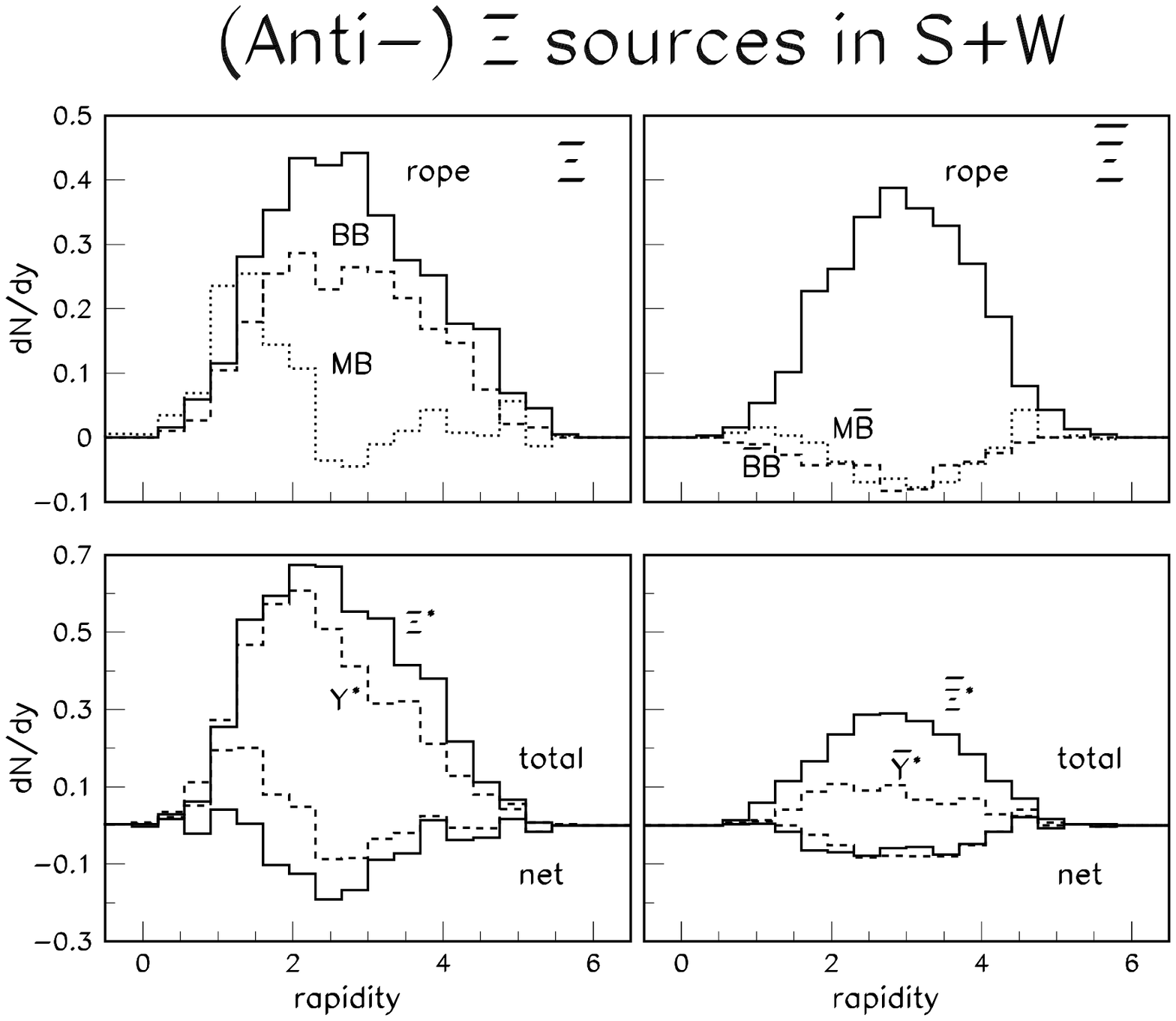,width=11cm,height=9cm}}}

\caption
[
 ]
{
 \label{swcoll}
 \underline{Sources of  $\Xi$ and  $\overline{\Xi}$  
    in S(200AGeV) on W from RQMD  \cite{SOR95PLB} }
}
\end{figure}

  If we want to shed more light on  the mechanisms which are responsible
   for the observed strange anti-baryon enhancement we have to go
   to the smaller systems. An analysis of chemical equilibration
   in the  strange (anti-)baryon sector for S(200AGeV) on W
   collisions has been made  in  \cite{SOR95PLB}.
  The main result can be extracted from Fig.~\ref{swcoll} in
  which the total and the net contributions to   $\Xi$ and  $\overline{\Xi}$ 
  are displayed. 
   The upper part shows the important net contributions of the different
    production mechanisms ($MB$= meson-baryon collisions and so on).
   The lower part breaks the $MB$ contributions up into  two parts,
    from intermediate states with cascade ($\Xi ^\star $), respectively
    hyperon ($Y  ^\star $) quantum numbers.
   It is clear that reactions with these intermediate quantum numbers
   are responsible to equilibrate hyperon and cascade states, e.g.\ 
      $\Lambda \bar{K} \longleftrightarrow \Xi \pi$ (and analogously for
    the anti-particles).  
   Even in  this light ion induced reaction the back and forth between
   the different baryon states
   seems to be rather strong.
  It is apparent that the total contributions in the central rapidity region
  are considerably larger than the net contributions. 
   This  means in turn that
  these strange baryon species 
   must be locally rather close to chemical equilibrium values. 
  
  The microscopic calculations with RQMD provide some justification for 
   the equilibrium assumptions underlying
    thermal fireball fits. An example of these fits is taken from
    Cleyman's paper   \cite{JCLE97}  and shown in Fig.~\ref{wa85fb}.
   One of the truely intriguing aspects is that the temperature 
   extracted from the fit of strange (anti-)baryons is very large,
   close to 190 MeV. This tells immediately that $T$ cannot be a real
   freeze-out temperature of a hadronic gas. 
    A hadron density of approximately  1 fm$^{-3}$ is  so high  that particles 
    cannot decouple. 
    A more reasonable conclusion is that the yields of these species 
   do not change appreciably
   any more below  this temperature (chemical freeze-out).
   According to the transport calculation with RQMD 
    and other models \cite{JCLE97} early chemical freeze-out
     is caused by the small hadronic strangeness creation
    and absorption rates. The relevant cross sections 
     are typically just a couple of
    millibarns. 
    The range of extracted chemical freeze-out temperatures 
   is  very close  to -- maybe even above --
    the phase transition
    temperature to the QGP. However, no  QGP  phase transition 
    is incorporated into RQMD. 
   The good agreement of RQMD with strangeness measurements could be an
    indication  that the properties of hadronic matter change very 
    smoothly into QGP properties in the transition region.
    
  Sometimes -- as in  Fig.~\ref{wa85fb}  -- an extra parameter
  is introduced to account for suppression  of hadrons
   containing strange valence quarks compared to equilibrium values.
  This strangeness suppression parameter is 0.7 in  Cleyman's fit,
   roughly a factor of 2 larger than in corresponding fits  to
    particle yields from $pp$ or $e^+$$e^-$ interactions.
  It seems to indicate that the
   systems created in $S$ induced reactions are much closer to  chemical
   equilibrium than the small systems. 
  A very important question in this respect is whether there is some
  {\em threshold} effect for strangeness enhancement as a function of
   system size. 
  Clearly, strangeness  production depends on the energy density achieved.
  Does it also depend on whether the size of the production volume 
 exceeds the confinement scale of one fermi?   Does it 
 depend on the degree of thermalization which is achieved in these reactions?
  These questions are of uttermost importance in the context of strangeness
  as a signature for QGP in the thermodynamic sense. 
  The place to look for  answers to these questions is
  $p$$A$ collisions, because the fire cylinder in these reactions has 
  a small transverse size  given roughly by the 
  radius of the in-going projectile proton.
  Furthermore, the life time of the system created in $p$+$A$ collisions
  is expected to be rather small which works against equilibration.
   
\par         
\begin{floatingfigure}{8.5cm} 

\mbox{
\epsfig{figure=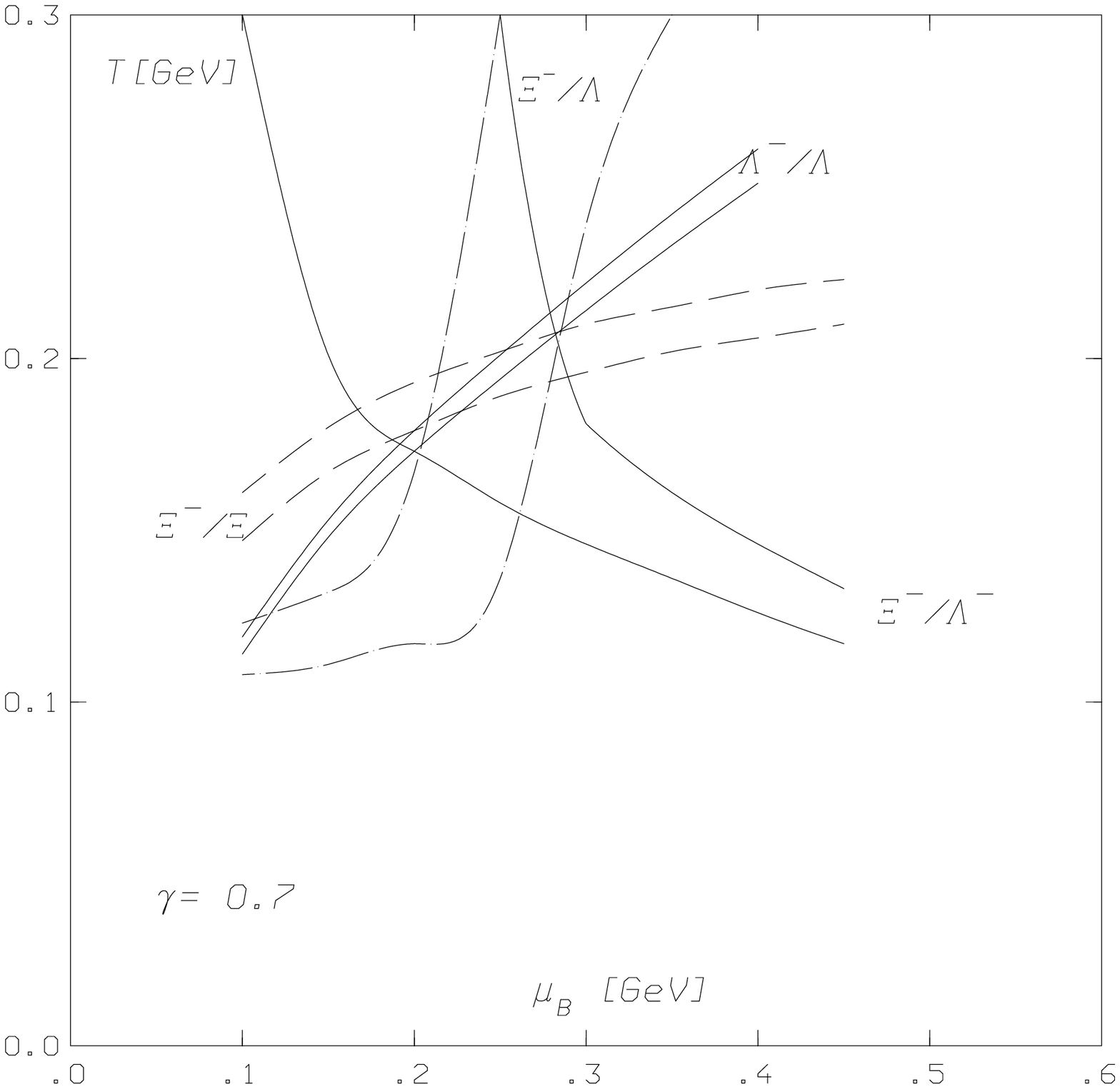,height=8cm,width=8cm}}   
  \caption[
          ]
                    {
 \label{wa85fb}
  {\small
  \underline{
 Fireball fit to 
  S(200AGeV) on 
  W  \cite{JCLE97} 
    }}
                            }
 \end{floatingfigure}
   \quad              
  Recently, it has been fiercely debated whether 
   $p$ collisions with nuclei as small as $S$ show `strangeness enhancement'
    or not  \cite{POP95,GAZ96}. 
   The answer depends somewhat on the definition.              
   Integrated $K$/$\pi $ seems not much affected by the presence of nuclei.
   Some $K^+$ enhancement is seen at backward rapidities    
    which is expected from rescattering effects leading to associated
   production  \cite{SOR93ZFP}. This is consistent with the observation that
    hyperons scale non-linearly with mass number.
   The most remarkable $p$$A$ result concerning strangeness has been published
   only recently. 
  The   multiple     strange antibaryon yields   
  reveal  that   production volume and  degree of thermalization are not
    relevant variables for their production. 
    WA94 finds that the 
   $\bar{\Xi}$/$\bar{\Lambda}$ value already in p(200GeV) on S 
   (and W) is intermediate between
  the $p$$p$ and the $S$+$S$/$W$ results \cite{WA94plb}. This means that
   two overlapping strings as typically produced
  in $p$+$S$ collisions are sufficient to create a strong departure
  from particle ratios in $p$$p$ collisions.
 It was already discussed in  \cite{SOR93ZFP} that ropes affect strongly the
   strange anti-baryon yields in  $p$+$A$  reactions.
   String fusion to color ropes provides an     
  example in which strangeness creation is  sensitive  to the
   local energy densities but not to the  production volume. 
   Whatever the best effective theory to describe $p$$A$ and $A$$A$ reactions
    one conclusion can be safely drawn from the experimental data alone.
   Strange hadron yields 
    per pion change smoothly from  $p$$p$ to $p$$A$ and $A$$A$.
   No clear threshold behaviour is observed.      
  This is bad news and good news at the same time.
   The question whether a 
   quark-gluon plasma state 
   of {\em macroscopic} dimensions in space and time 
   is created in $A$$A$ collisions can probably not be answered by looking at
  the strange hadrons. 
   On the other side, all information on   strange hadrons 
    produced in the small systems point to
   strong influence of  variations in energy density on the
    particle ratios. 
   The observed leveling-off of  strange particle ratio  like 
      $\bar{\Xi}$/$\bar{\Lambda}$ 
    in  data taken with the $S$ and the $Pb$ beam \cite{GODY97}
    tells that the truely heavy-ion collisions do not probe so much
    the production mechanisms any more but the subsequent equilibration
    stages.
   An experimental program with $p$ and very light ion beams 
   at the CERN-SPS energies is required
  if one would like to study strangeness production mechanisms carefully.
  This is also warranted in view of present experimental uncertainties 
   and conflicting data sets, even on the level of $p$$p$ collisions. 
  Another direction to proceed would be to study  heavy-ion collisions
   at lower  beam energies. Most interestingly, 
    $\bar{\Lambda}$/$\bar{p}$ seems to be
   much larger at the  beam energy of 10~AGeV than at 200~AGeV.
    What about the ratio  $\bar{\Xi}$/$\bar{\Lambda}$  
    in Au on Au at 10~AGeV? Nobody knows the answer.
   There must be a lower beam energy at which the  inversion
    that  heavier anti-baryons are populated more frequently than
    the lower ones goes away due to lack of phase space.  
    Enhancement (and absorption) effects are expected to be much more
    pronounced at lower beam energies which should help to clarify
    their dynamical origins.
   
\par         
\begin{floatingfigure}{8cm} 

\mbox{
\epsfig{figure=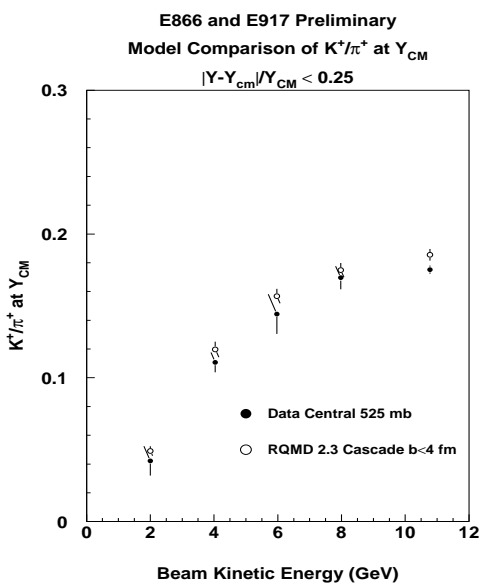,height=8cm,width=8cm}}   
  \caption[
          ]
                    {
 \label{kpi_exc}
  {\small
  \underline{
     $K$/$\pi$ in Au+Au, 2-12~AGeV    
    \cite{DUN97} 
    }}
                            }
 \end{floatingfigure}
   \quad              
  So far I have focused on equlibration in the strange (anti-)baryon sector.
  Of course, the kaon degrees of freedom should get equilibrated
  much more easily because of the comparably smaller masses. 
 Arguments  have even been put forward  that the   $K$/$\pi$ ratios 
   measured in $A$$A$ collisions at CERN energies  
 signal the  QGP formation. Some authors argue that the   $K$/$\pi$ ratio
  is anomalously large  \cite{GAZ96} which is based on a comparison
  to $p$$p$ data. On the other side, it is also sometimes  suggested
   (even  by  the same  authors)  that the ratio 
   is  anomalously  small   \cite{LET95}. 
  This is based on the argument 
    that the hadronic fireball
  fits  work well in the strange (anti-)baryon sector but predict
   a too large  $K$/$\pi$ ratio \cite{DAV91} in $S$ induced collisions.
   Indeed, the observation of a `pion excess' indicates an
     interesting non-equilibrium effect. 
    A simple reason has been
   extracted from the RQMD transport calculations as was pointed out
   in  \cite{SOR95PLB}. The source size of pions is systematically larger
   than the source of kaons and (anti-)baryons. On the one side,
  this is a consequence
   of the different production mechanisms. For instance, anti-baryons
   are produced more centrally in the impact parameter plane than pions,
    because their formation rate is more sensitive to the higher
   energy densities in the interior (the overlapping strings
   and formed ropes in the RQMD model).
   Furthermore, the pion `fluid' expands considerably faster than the
   heavy hadron fluid  \cite{sorplb96}.
   A comparison of RQMD  predictions with recent preliminary NA49 data
    for central Pb(158AGeV) on Pb collisions
   \cite{SOR97PLB} establishes that this type of {\em non-ideal}
   hydrodynamic expansion is essential  for the good agreement between
   the shape of the calculated transverse mass spectra  and data.
   Note also that 
   RQMD has predicted not just the shape but also
   the measured  $K_S$/$\pi$ ratio  correctly.  
   
  A   test of the idea  whether the  $K$/$\pi$ ratio 
  can be linked to the QGP formation is to lower the beam energy and
  look for a  discontinuity in the excitation function. Such a
   test has been realized recently at the BNL-AGS by varying the beam
   energy between 2 and 12~AGeV (see C.\ Ogilvie's talk).
   The data are still very preliminary and are shown in
   Fig.~\ref{kpi_exc}. They are compared in this Fig.~with the predictions
   from the RQMD model. The agreement is very good. Neither data nor model
   show any discontinuity in this variable. 
   
   One can go back to the model and look for the dominating source
    of strangeness enhancement \cite{SOR91}. 
   It turns out that the crucial mechanism is 
    the hard collison spectrum in the pre-equilibrium stage leading to    
    the formation of  excitations (resonances or strings), predominantly
    via meson-baryon interactions. Since the resonances are progated
    themselves even larger mass excitations may be created 
   in multi-step processes. 
   
 These processes, e.g.
\begin{displaymath}      
  \hspace{12em}
\begin{array}{ccccc}
  \pi + N &          &       &          &  \pi +  \pi  \\
          & \searrow &       & \swarrow &              \\   
          &    &  \Delta + \varrho &    &              \\   
          &       &  \downarrow  &      &              \\   
          &       &\Delta ^\star &      &              \\   
\end{array}      
\end{displaymath}    
   are of huge importance for strangeness creation, because
  intermediate   resonances  act as an `energy storage'.
  Such multi-body hadronic interactions are only frequent in
  a system of sufficient density of roughly 1 fm$^{-3}$,
   because the life time of the intermediate states amounts to
   typically 1-2 fm/c only. 
   
  Resonances, in particular the $\Delta $(1232), impact          
  the dynamics in $A$$A$ collisions  already 
  at beam energies of  1-2~AGeV \cite{MET97}.
  Going to higher beam energies means at first that more and more resonance
  states can be excited. This is reflected in the various attempts
  for a theoretical description of $A$$A$ collisions in the region
  of beam energies above 1~AGeV.
  A transport model like RQMD contains a whole
  tower of resonance excitations, basically all confirmed resonance
  states up to a mass of 2 GeV (including the missing  members of 
  $SU(3)_F$ multiplets required by the quark model). 
  {\em  Resonance matter} formation  seems to be   a crucial aspect of
   the physics of hot, dense hadronic matter.                
   The resonance spectrum is related primarily to
  the confinement property of QCD. 
  In equilibrium, 
   the excitation of the resonance states
    with their large degeneracy factors leads to a strong increase of
    energy (and entropy)
   density within a small temperature range 150-170 MeV.
    The most extreme variant of such an approach 
   is the Hagedorn gas with an exponentially
    increasing spectrum of states as a function of
 mass which leads to a limiting
    temperature.
  Is  resonance matter  just a way of describing the
    `mixed phase' in the transition region employing hadronic degrees
   of freedom?               
 In fact, excitation of  resonances provides even a path
 to chiral symmetry restoration.    Glozman and Riska  argue that 
  the  splitting between parity doublets  which characterize the amount of
   spontaneous breaking of chiral symmetry
   goes to zero for higher lying flavor  multiplets  \cite{GlRi96}.  
  It is an open and interesting question how far one can push the idea of
 {\em duality} between descriptions on the hadronic and the quark-gluon level
  in the transition region.
 Another interesting idea is that hadron properties -- e.g.\ masses and
 decay widths -- themselves might
  change in the dense medium. Since these properties are tied to the
  spontaneous breaking of chiral symmetry, partial restoration of
   this symmetry at finite density and temperature may lead to these
  changes. Furthermore, effective theories 
   like chiral perturbation theory
   predict medium effects --
   even without chiral restoration. 
  Studies of strangeness might help
 to shed some light on this question. Does subthreshold production of
  $K^- $ in $A$$A$ collisions at GSI-SIS energy give an indication
  for a dropping  $K^- $ mass? This is an exciting possibility which
  certainly deserves thorough experimental and theoretical studies
  (see G.Q.~Li's talk, \cite{GQLi97}).

\section{Strangeness and the post-equilibrium stage} 
   
\begin{figure}[tb]

\centerline{\hbox{
\psfig{figure=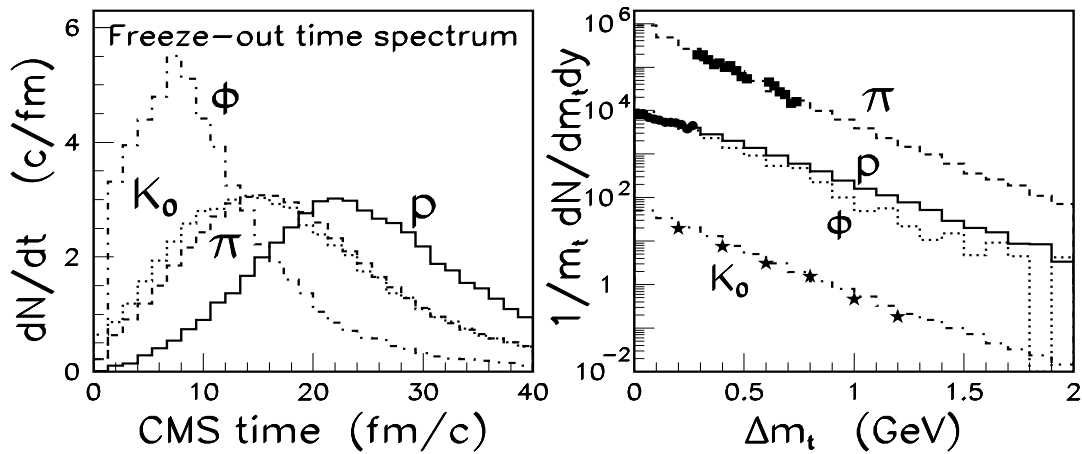,width=11cm,height=6cm}}}
  \caption[
          ]
                    {
  {\small
 \label{pbpb_phi}
   Freeze-out time and transverse mass ($m_t$)
    spectra of hadrons at midrapidity
    calculated with RQMD 2.1.
    $m_t$ spectra are compared to preliminary NA49 data.
    Yields have been renormalized for  display reason.
    }
                            }
\end{figure}

 The  post-equilibrium stage develops in $A$$A$ collisions 
  when the reaction rates are not large enough to 
  keep the expanding  system in equilibrium. 
  (Of course, it is  possible that local equilibrium has not been
   achieved at all  during  intermediate stages.)
  The criterion from kinetic theory is that the associated   
   relaxation time becomes larger than the expansion time scale.
 In particular,  in a relatively baryon-number poor environment as created at
   CERN energies
  this is going to happen  quickly for the rare anti-baryons.
   Their absolute number  probably freezes-out rather early. 
   Lateron  the strangeness content  goes out of equilibrium,
    because it changes too slowly.
   The effective pion number (counting rho as two pions and so on) is
   freezing-out approximately at the same time as strangeness. 
  The reason is  that
   reactions of the type $\pi\pi  \pi  \pi  \leftrightarrow \pi  \pi  $
    are  characterized by   small cross sections, too. Furthermore,
    pion number may change only by at least two  units in elementary 
   reactions in the meson sector (from $G$ parity). 
   Kinetic equilibrium in the expanding hadron gas 
  can probably be maintained for a longer time.  No kinematic threshold 
   prevents the occurence of elastic reactions. Since
   hadrons with strange valence quark content have typically smaller
   elastic or quasi-elastic cross sections, their interaction tends to 
  cease earlier.
   Another simple rule based on quark counting 
   and empirical observation is that mesons have usually smaller interaction
   rates than baryons.
   Of course, the Goldstone theorems makes sure that pion interactions
   at low energies are dynamically suppressed. 
   Therefore pions will decouple earlier than for instance omega mesons
   or nucleons.
   Transport calculations have been analyzed to study the 
    post-equilibrium effects
    very thoroughly  \cite{BRA95,sorplb96}. 
    The  spectrum of freeze-out times for different hadron species
    produced at midrapidity in central Pb(160AGeV) on Pb collisions
     according to RQMD calculations 
     is displayed in Fig.~\ref{pbpb_phi}.
    It confirms the stated trends. Note that the double-strange $\Phi $ mesons
     decouple from the system about 12 fm/c earlier than the nucleons.
   
    An impressing body of evidence has been accumulated in recent years
   that  models like RQMD, VENUS and ARC produce a final source
    which is a very good approximation of the real source in 
     phase (position and momentum) space. In this respect
     HBT analysis (see W.~Zaic's talk) 
      and nuclear cluster calculations 
     were very important, because  the space-time characteristics
     of the  sources as calculated in  these models
       is completely determined by the interactions. 
   The realistic simulation of the
   final dilute stages in nucleus-nucleus reactions 
    -- including the  break-down of equilibria -- 
   appears to be
    an important reason why the microscopic models are so
   successful compared to experimental data. 
   This is most clearly revealed by comparison to hydrodynamic calculations
   which usually assume the existence of local equilibrium
     until the final freeze-out and therefore lack the 
    sequential break-down of equilibria. For instance, 
    momentum distributions  calculated  with a hydrodynamical code 
     for $S$ induced reactions at 200~AGeV are
    presented and discussed in \cite{SOL97}.
   It is apparent from the comparison to experimental data that sequential
    freeze-out could considerably improve the agreement. In this particular
   calculation a `compromise' value for the freeze-out temperature
    has been chosen. This means that strange anti-baryons like
    $\bar{\Lambda}$ are underpredicted, because the chosen temperature is
   already too `cool'. On the other side, the hadron spectra are generally
   too soft  compared to  the measurements indicating that the system
    freezes-out too early:  the transverse flow has not yet  fully
   developed.
   Presently, several groups are studying how 
    the freeze-out stage can be better simulated  either intgrated into or as
   an `after-burner' to the hydrodynamics calculations.
   
  The importance of the post-equilibrium stage increases strongly
  with the masses of projectile and target. 
  As pointed out in \cite{SOR95}  
  the source which is produced in Pb+Pb reactions
  lives much longer  
  in the hadronic stage than the source created e.g.\ in S+Pb. Three-dimensional
  expansion becomes effective only much later in the heavier system. 
  The reason is that
  the initial transverse size of the reactione zone is larger by
  almost a factor of four.
  (In the limit of infinite transverse size expansion would be purely
  longitudinal.)
  Furthermore, the larger system cools down to lower kinetic temperatures
   \cite{SOR95PLB}, because the  initial size affects  also the 
   decoupling conditions.

  Let me give just one example how the sequential decoupling of
  hadron species from the medium can be utilized to gain information
  on properties of the medium. Flow analysis is
   one of the most promising areas to study the physics of the phase
  transition.  Expansion becomes very `soft' in the transition region
     affecting the flow of the matter. 
   There are many speculations whether the observed rather hard transverse
   momentum spectra in Pb(160AGeV)+Pb reactions  
    -- slope parameters  close to 290 MeV for (anti-)baryons -- 
   reflect an `early'
   (partonic?) flow  or `random kick' \cite{Leo96} 
  or -- alternatively -- develop only very late in the 
   hadronic stage.
   A comparison of $\Phi $ and $p$ spectra is very useful, because these
   two species have practically the same mass. 
   The random $p_t$ component which they acquire is therefore the same
   in a thermal (or hydrodynamical) framework. RQMD predicts that
   the $\Phi $ spectrum is in fact considerably softer (slope 225 MeV)
 than the $p$ spectrum with slope 280 MeV (see the rhs of Fig.~\ref{pbpb_phi}).
  This seems to indicate that the additional flow seen in Pb+Pb
   (beyond what has been observed in p+A and S+A at 200~AGeV) is
   produced rather late only, after the $\Phi $ has already decoupled.
  The results support the idea 
     that initially
   the source at CERN energies expands very softly into transverse directions.
  It is another (interesting) question whether this is a pre-equilibrium
   effect (e.g.\ from ropes) or manifestation of the `softest point'
   in the Equation of State \cite{SHU97}.
   A general lesson to be drawn from recent experience gained with
    $A$$A$ reactions at the ultra-high but also at the relativistic
   (1~AGeV) energies  \cite{GQLi95}  is that strange hadrons 
    are a very valuable tool to disentangle the various contributions
   to the  forces which drive the  collective  flow.

\section{Conclusions} 
   
  I find evidence from the systematics of measurements
  and theoretical studies that strangeness enhancement per se is not
   a signature of  QGP formation in ultrarelativistic
   nucleus-nucleus collisions.
  On the other side, 
  strange
 anti-baryon enhancement at 160~AGeV -- and probably even more so at
  10 AGeV  --  signals importance of  interactions beyond    
 (quasi-free) hadron gas dynamics. 
 If one wants to study `threshold behaviour' carefully one has to go back
  to small  systems like  $p+A$  or to lower beam energies.
  Concerning the heavy-ion reactions 
 emphasis has shifted recently from study of production mechanisms to
 equilibrium and post-equilibrium  physics. 
 This reflects that  strange hadrons -- even the multiple strange
 anti-baryons -- are produced frequently enough to 
  achieve a large degree of chemical
 equilibration in the center of the `fireball'.

This work has been
supported by DOE grant No. DE-FG02-88ER40388.


\begin{thebibliography}{999}
   
\bibitem{KOC86}
 P.\ Koch, J.\ Rafelski, and B.\ M\"uller:
 {\em Phys.\ Rep.} {    142} (1986)  167.
   
\bibitem{NA3590}
J.\ Bartke et al. (NA35 Collab.):
    {\em Z.\ f.\ Phys.} {    C48} (1990) 191; 
J.\ B\"achler et al. (NA35 Collab.):
    {\em Z.\ f.\ Phys.} {    C61} (1994) 551.    
   
\bibitem{ABA91}
 S.\ Abatzis  et al. (WA85 Collab.):
 {\em Phys. Lett.} {    B244} (1990) 130;
 {\em Phys. Lett.} {    B259} (1990) 508;
 {\em Phys. Lett.} {    B270} (1991) 123. 
   
\bibitem{HIPAGS96}
J.G.\ Lajoie  (for the E864 Collab.):
Proceedings of the
Workshop ``Heavy Ion Physics at the AGS'' (HIPAGS '96), 
 Wayne State University 1996, WSU-NP-96-16.
   
\bibitem{BMSWX96}   
 P.\ Braun-Munzinger, J.\ Stachel, J.P.\ Wessels,
 and N.\ Xu:
  {\em Phys.\  Lett.} {    B344} (1995)  43; 
   {\em Phys.\  Lett.} {    B365 (1996) 1.
   
\bibitem{SBM96}   
 P.\ Braun-Munzinger, J.\ Stachel: 
      {\em  Nucl.\ Phys.} {    A606} (1996)  320.
   
\bibitem{BEB92}
H.\ Bebie, P.\ Gerber, J.L.\ Goity, and H.\ Leutwyler:
      {\em  Nucl.\ Phys.} {    B378} (1992)  95.
   
\bibitem{SOR97PLB}
H.\ Sorge:
 SUNY-NTG 97-1, nucl-th/9701012,
    Phys.  Lett. B  (1997) in print.           
   
\bibitem{SOR95ZFP}
  H.\ Sorge:
     {\em Z.\ f.\ Phys.} {    C67} (1995) 479. 
   
\bibitem{SOR92}
     H.\ Sorge, M.\ Berenguer,
             H.\ St\"ocker, and W.\ Greiner:
            {\em Phys.\ Lett.} {    B289} (1992) 6.

\bibitem{ABFP95}
 N.\ Armesto, M.A.\ Braun, E.G.\ Ferreiro, and C.\ Pajares:
            {\em Phys.\ Lett.} {    B344} (1995) 301.  
   
\bibitem{CBOR97}
C.\ Bormann (for the NA49 Collab.):
 Contrib.\ to the Proceedings
 `Strangeness In Quark Matter 1997',  Santorini
 (Crete) April 1997.
   
\bibitem{SOR95}
 H.\ Sorge:
  {\em Phys.\ Rev.} {    C52} (1995) 3291.  
                                                      
\bibitem{CAP95}
 A.\ Capella: 
    {\em Phys.\ Lett.} {    B364} (1995) 175.  

\bibitem{WER93}
 K.\ Werner:
    {\em  Phys.\ Rep.} {    232}  (1993)  87.    
   
\bibitem{GEI97}
 K.\ Geiger, D.K.\ Srivastava:
  nucl-th/9706002,
 K.\ Geiger, R.\ Longacre:  nucl-th/9705041.

\bibitem{BLZ95}
T.S.\ Biro, P.\ Levai, and J.\ Zimanyi:
 {\em Phys. Lett.} {    B347 } (1995)  6.

\bibitem{RLT97}
 J.\ Rafelski, J.\ Letessier,  and A.\ Tounsi:  
 {\em Phys. Lett.} {    B390} (1997)  363.
   
\bibitem{JAH94}
  A.\ Jahns et al.:
 {\em Phys.\ Rev.\  Lett.} {    72}  (1994) 3464.  
   
\bibitem{ARC93}
 S.H.\ Kahana, Y.\ Pang, T.\ Schlagel, and C.B.\ Dover:
      {\em Phys.\ Rev.} {    C47} (1993) R1356. 

\bibitem{SOR95PLB}
  H.\ Sorge:
   {\em Phys.\  Lett.}  {    B344 } (1995) 35.  
   
\bibitem{JCLE97}
J.\ Cleymans:
 Talk presented at the `Third International Conference on
 Physics and Astrophysics of Quark-Gluon Plasma',
 March 1997, Jaipur, India, nucl-th/9704046.

\bibitem{BRA95}
L.V.\ Bravina, I.N.\ Mishustin, N.S.\ Amelin,
J.P.\ Bondorf, and L.P.\ Csernai:
  {\em Phys.\  Lett.} {    B354} (1995) 196.  

\bibitem{POP95}     
V.\ Topor Pop et al.:
  {\em Phys.\ Rev.} {    C52} (1995) 1618;
 M.\ Gyulassy, V.\ Topor Pop, and X.N.\ Wang:
  {\em Phys.\ Rev.} {    C54} (1996)  1497.
   
\bibitem{GAZ96}
M.\ Gazdzicki, U.\ Heinz:
  {\em Phys.\ Rev.} {    C54} (1996)  1486.
   
\bibitem{SOR93ZFP}
H.\ Sorge, L.\ Winckelmann, H.\ St\"ocker, and W.\ Greiner:
     {\em Z.\ f.\ Phys.} {    C59} (1993) 85.
                                                  
\bibitem{WA94plb}
 S.\ Abatzis  et al. (WA85 Collab.):
 {\em Phys. Lett.} {    B2400 (1997) 239. 
   
\bibitem{GODY97}
G.\ Odyniec (for the NA49 Collab.):
 Contrib.\ to the Proceedings
 `Strangeness In Quark Matter 1997',  Santorini
 (Crete) April 1997.
   
\bibitem{DUN97}
 Courtesy  of  C.\ Ogilvie
 (see his contribution to the proceedings)
  and  J.\ Dunlop 
  who made the comparison with RQMD.

\bibitem{LET95}
 J.\ Letessier, A.\ Tounsi, U.\ Heinz, J.\ Sollfrank, and J.\  Rafelski:
 {\em Phys.\ Rev.} {    D51} } (1995) 3408.
   M.\ Tazdzicki (for the NA35 Collab.):
      {\em  Nucl.\ Phys.} {    A590} (1995)  197c.       

\bibitem{DAV91}
 N.\ Davidson, H.\ Miller, R.\ Quick, and J.\ Cleymans:
          {\em Phys.\ Lett.} {    B255} (1991) 105.   
   
\bibitem{sorplb96}
 H.\ Sorge:
   Phys.\ Lett.\ {    B 373} (1996) 16.
   
\bibitem{SOR91}
 H.\ Sorge, R.\ Mattiello,
 H.\ St\"ocker, and W.\ Greiner:
 {\em Phys.\ Lett.} {    B271} (1991) 37.    

\bibitem{MET97}
See V.\ Metag's contribution.
                                         
\bibitem{GlRi96}
L.Ya.\ Glozman, D.O.\ Riska:
  {\em Phys.\ Rep.}} {     268}  (1996) 263.
   
\bibitem{GQLi97}
 G.Q.\ Li, C.-H.\ Lee, and G.E.\ Brown:
   nucl-th/9706057.  
   
\bibitem{SOL97}
J.\ Sollfrank, P.\ Huovinen, M.\ Kataja, P.V.\ Ruuskanen,
M.\ Prakash, and R.\ Venugopalan:
  {\em  Phys.\ Rev.}  {    C55} (1997) 392.
   
\bibitem{Leo96}  
 A.\ Leonidov, M.\ Nardi, and H.\ Satz:
      {\em  Nucl.\ Phys.} {    A610} (1996)  124c.       
   
\bibitem{SHU97}
E.\ Shuryak: hep-ph/9704449.
   
\bibitem{GQLi95}
 G.Q.\ Li, C.M.\ Ko, and B.A.\ Li:
 {\em Phys.\ Rev.\  Lett.} {    74}  (1995) 74.  
   
\end{thebibliography}
\end{document}